# Comparison of round- and square-core fibers for sensing, imaging and spectroscopy

MATTHIAS C. VELSINK,[1,2] ZHOUPING LYU,[2] PEPIJN W.H. PINKSE,[1] AND LYUBOV V. AMITONOVA[2,3*]

[1]*MESA+ Institute for Nanotechnology, University of Twente, PO Box 217, 7500 AE Enschede, The Netherlands*
[2]*Advanced Research Center for Nanolithography (ARCNL), Science Park 106, 1098 XG, Amsterdam, The Netherlands*
[3]*LaserLaB, Department of Physics and Astronomy, Vrije Universiteit Amsterdam, De Boelelaan 1081, 1081 HV Amsterdam, The Netherlands*
*l.amitonova@arcnl.nl*

**Abstract:** Multimode fibers (MMFs) show great promise as miniature probes for sensing, imaging and spectroscopy applications. Different parameters of the fibers, such as numerical aperture, refractive index profile and length, have been already optimized for better performance. Here we investigate the role of the core shape, in particular for wavefront shaping applications where a focus is formed at the output of the MMF. We demonstrate that in contrast to a conventional round-core MMF, a square-core design doesn't suffer from focus aberrations. Moreover, we find that how the interference pattern behind a square-core fiber decorrelates with the input frequency is largely independent of the input light coupling. Finally, we demonstrate that a square core shape provides an on-average uniform distribution of the output intensity, free from the input-output correlations seen in round fibers, showing great promise for imaging and spectroscopy applications.

## 1. Introduction

A probe made out of a piece of fiber optic is a key element for remote sensing, microscopy and spectroscopy as well as deep-tissue bioimaging [1]. Recently, conventional multimode fibers (MMFs) have turned into very promising tools for imaging [2,3] and spectroscopy [4] especially because of the development of wavefront-shaping techniques. The interference between the guided modes creates a distinct wavelength-dependent speckle pattern, thus providing a unique fingerprint of the input wavelength given the MMF configuration [5]. As a result, a standard MMF can be used as a general-purpose spectrometer. On the other hand, a single MMF can transmit many pixels of an image through its numerous propagating modes. Miniaturized fiber-based endo-microscopy methods offer a large penetration depth and are not limited by the interior of hollow organs. An MMF has a significantly higher mode density than the core density of a fiber bundle, leading to a higher spatial resolution combined with a smaller footprint. The numerical aperture (NA) of properly designed MMFs are comparable with conventional bulk imaging optics [6,7].

Multiple approaches have been developed to provide clear imaging through a single MMF probe: transmission matrix measurements alone [8] or in combination with holographic light control [2,9–11], time-domain wavefront shaping [12] and compressive imaging [13,14]. The most common techniques exploit *spatial* or *temporal (complex) wavefront shaping* to create a focal spot at the output fiber facet. The image is built by scanning of the focal spot on the output fiber facet. In theory, a perfect diffraction-limited spot can be created at any position at the fiber output facet. However, the limited number of controlled modes defines the quality of the foci. The distorted shape of the aberrated focal spot in multimode-fiber imaging with a low number of segments arises from the shape of the most contributing fiber modes [15]. Another scheme for MMF imaging by exploiting a sparsity constraint for robust image recovery has been proposed recently [13,14]. *Compressive imaging* through an MMF allows effective and fast image reconstruction by using only a subset of the transmission matrix and does not require complex wavefront shaping in space or time. It provides higher imaging speed and makes the setup simpler and more convenient for practical use. It relies on a nearly orthogonal set of uniformly distributed random speckle patterns generated by an MMF.

To summarize, most of the MMF-based imaging and spectroscopy methods have several common requirements: The need to excite and control many, preferable all, modes and the ability to cover the whole field-of-view (FOV) uniformly. Imaging methods exploiting only a single spatial mode of an MMF input are very important for the development of a flexible probe. Consequently, an optimized fiber should allow to excite many high-order modes via a single spatial input, providing uniform intensity distribution without aberrations on the output. Wavelength-dependent characterization of MMF-based endoscopes with different refractive-index profiles has been done without addressing the core shape, however [16]. Recently, new types of MMFs have been demonstrated [17]. Novel geometries for the core region encompass square and rectangular core cross-sections instead of the conventional round core. These shapes influence the modal structure of the propagating light and have a potential for improving the optical properties. A rectangular-core MMF can be used to match rectangular on-chip waveguides for wavelength demultiplexing [18]. On the other hand, conservation of the wave vector in a short rigid square-core MMF provides new opportunities for nonlinear imaging [19].

Here we explore the influence of the shape of the core on imaging and on the spectral sensitivity of MMFs. We compare two experimentally most-relevant core shapes: a round-core MMF, which is the most common type, and a square-core fiber. The latter is expected to provide better ray mixing and, consequently, a flatter, top-hat-like intensity profile [20]. We demonstrate that MMF imaging with a square-core fiber doesn't suffer from the aberrations known from round-core MMFs and provides a more uniform distribution of the speckle-like output intensity, showing great promise for imaging, sensing and spectroscopy applications.

## 2. Experimental setup

Our experiments are performed on step-index multimode fibers with an NA = 0.22, a length of 1 m and different core shapes: a round-core fiber with a diameter of 105 μm (Thorlabs) and a square-core fiber with a core size of 70x70 μm$^2$ (CeramOptec). The fibers were coiled up with a bend radius of about 10 cm. The microphotographs of the square- and round-core MMFs are shown in Fig. 1. The square-core fiber has rounded edges with a radius of curvature of about 5 μm due to the limits of the fabrication process. However, this is not critical for the mode structure. The round-core fiber sustains approximately $N_{modes} = V^{\,2}/2$ modes [21], where $V$ is the V-number or normalized frequency of the fiber. It can be calculated as $V = 2\pi\, a\, NA/\lambda$, where $a$ is the core radius, $\lambda$ is the wavelength. For a wavelength $\lambda$ = 532 nm our round-core fiber supports approximately 9000 modes. To estimate the total number of modes supported by the square-core fiber, we use the following approximation: $N_{modes} = kV^{\,2}/2$, where the ratio $k = S_{square}/S_{round}$, and $S_s$ is the area of the fiber core with shape s. For a wavelength $\lambda$ = 532 nm our square-core fiber supports approximately 5400 modes.

For wavefront-shaping experiments, we use the second-harmonic output of a continuous-wave Nd:YAG laser with a wavelength of 532 nm (Cobolt Samba). A sketch of the experimental setup is presented in Fig. 1(a). A half-wave plate and a polarization beam splitter are used to adjust the laser power. The beam is expanded by telescope (L1 f1 = 50 mm, L2 f2 = 150 mm) to match the area of digital micromirror device (DMD) from Texas Instrument (1920×1200 tilting micromirrors) driven by the DLP V-9500 VIS module (Vialux) which is used to perform phase control of the beam. The complex wavefront shaping has been done using a stepwise sequential algorithm [22] and the Lee holography method [23]. The mirrors of the DMD were used to create a binary 2D grating. The DMD area illuminated by the beam is divided into $N_{seg}$ segments. A 4f configuration (L3 f3 = 150 mm, L4 f4 = 100 mm) images the DMD on the back focal plane of the first objective lens (Obj1). The pinhole (P2) blocks all the diffraction orders but the +1, which encodes the desired spatial phase distribution. Objective #1 (Olympus, 20×, NA = 0.4) couples the beam to the MMF and Objective #2 (Leica, 40×,

NA = 0.6) focusses the output light on the camera. We individually modulate the phase of each segment by spatially shifting the corresponding grating pattern between 0 and $2\pi$ in three steps of $2\pi/3$ each. For each modulated segment, the phase leading to the highest intensity in each spot on the fiber output is calculated by fitting a cosine function over three measurement points. By repeating the measurement for each DMD segment, the optimal phase pattern is reconstructed and the intensity on each target spot can be enhanced relative to the uncontrolled initial speckle pattern.

For a wavelength-scanning experiment, we use the continuous-wave linearly polarized output of a tunable Ti:Sapphire laser (Coherent MBR). The setup is presented in Fig. 1(b). A half-wave plate and a polarization beam splitter are used to adjust the laser power. The wavelength of the laser is manually tuned, and measured with a wavemeter (Burleigh WA-10L). Objectives #3 and #4 (Olympus, 20×, NA = 0.4) are used to couple light in and out of the fiber, respectively.

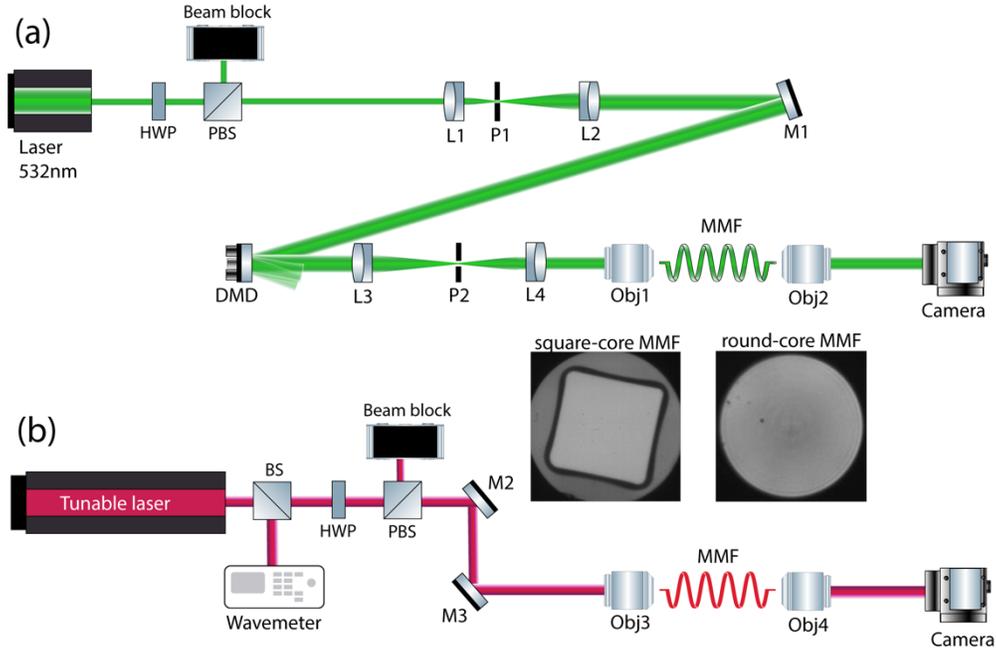

Fig. 1. (a) Experimental setup for the spatial-domain wavefront shaping through an MMF. (b) Experimental setup for the wavelength dependence measurements. The microphotographs of the square- and round-core MMFs are shown as an inset. MMF, multimode fiber; DMD, digital micromirror device; PBS, polarizing beam splitter; BS, beam splitter; HWP, half-wave plate; M, mirror; L, lenses; Obj, objectives; P, pinhole.

## 4. Results

### 4.1 Spatial wavefront-shaping approach

The most popular approach of MMF imaging requires transmission matrix measurements in combination with wavefront shaping – holographic control of light at the fiber input. The imaging quality is determined by our ability to create diffraction-limited foci over the full field of view (FOV) from the center toward the edges. In this section, we compare the distortion of the resulting foci for a round-core and a square-core MMF.

In the first set of experiments, a total of 16 × 16 foci have been independently optimized at different positions on the distal end of the round-core MMF as described above. The foci have been organized in a square grid with a spacing between two neighboring foci of about 3 μm. After a single optimization loop for a fixed number of controlled segments on the DMD, $N_{seg}$, we sequentially display every optimized phase pattern on the DMD and image the output intensity picture with the camera. Images of all foci have been recorded with the same acquisition time of the camera. To demonstrate the quality of each focal spot, we display the incoherent sum of all recorded pictures in Fig. 2(a) for $N_{seg}$ = 441 (left), 1225 (center), 1936 (right). We see that in the case of a round-core fiber the focus becomes elliptical, with the smaller axis directed toward the center of the fiber, as the focus position moves toward the fiber's cladding. While the quality of the focal spots increases with the number of controlled segments, foci close to the edge have an elliptical shape, even for very high $N_{seg}$. We also see that the power degrades toward the edges of the round-shaped core.

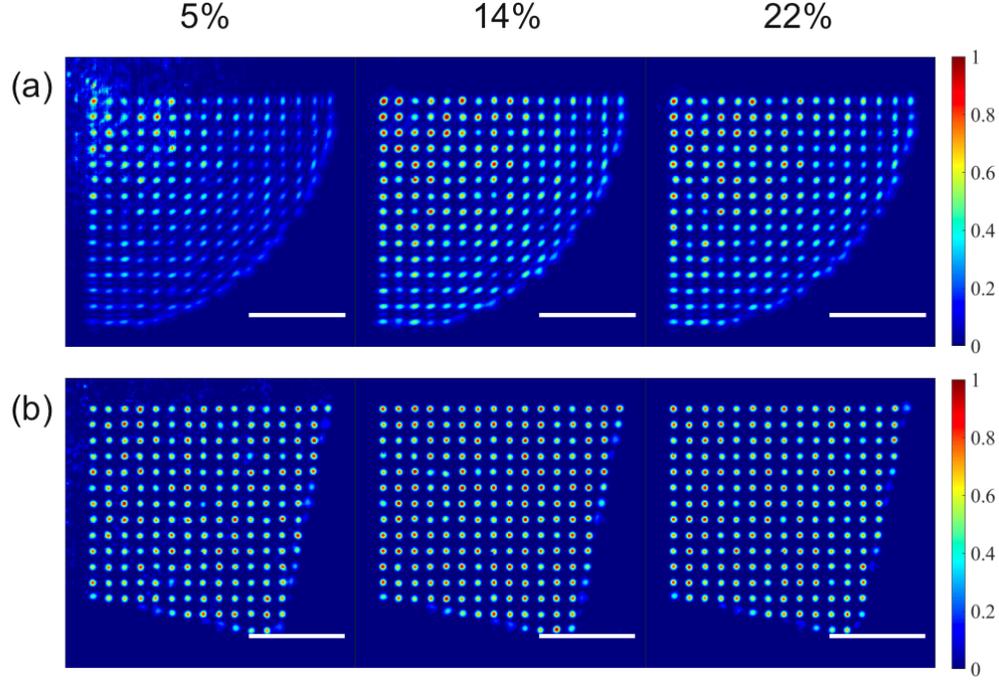

Fig. 2. Mosaic from the intensity images of 16 × 16 independently optimized foci at the distal end of the MMFs with a round core (a) and a square core (b). For wavefront shaping $N_{seg}$ = 5% (left), 14% (center), 22% (right) of the total number of modes supported by the corresponding fiber has been used. The scale bars are 20 μm and each incoherent sum has been normalized to the highest intensity.

In the second set of experiments, the same square grid of 16 × 16 foci with about 3 μm spacing has been independently optimized on the distal end of the square-core MMF. The mosaic built from all recorded pictures is presented in Fig. 2(b) for $N_{seg}$ = 256 (left), 676 (center), 1024 (right). The number of segments was chosen to maintain the ratio between the number of segments and the total number of modes at the level of 5%, 14% and 22%, respectively. Even for a relatively low number of segments, the intensity distribution is more uniform and all the foci have a symmetric round shape.

For a quantitative analysis of the better uniformity and focus shape of the square MMFs seen in Fig. 2, we characterize each focal spot. We measure the full width at half maximum (FWHM) of the foci located at the different positions from the core edge. The FWHM focal dimensions in the tangential and axial directions are experimentally measured as functions of the distance from the core edge (where the edge position is defined as 0) for the round-core fiber and are presented in Fig. 3(a) and 3(b), respectively. Different colors represent the different number of DMD segments used for wavefront shaping: $N_{seg}$ = 441 (blue), 1225 (red), 1936 (yellow). The solid green line indicates the FWHM that corresponds to the diffraction limit of the round-core MMF, calculated as $\lambda/(2NA)$, where $\lambda = 532$ nm and NA = 0.22. Fig. 3(a) shows that the round-core fiber suffers from aberrations in an about 25 μm-wide area close to the edge. The focus quality is slowly improving with an increasing number of segments. However, if the fiber supports a large number of modes (9000 in our case), it is practically very hard to control enough segments to overcome these aberrations. We also characterize the intensity distribution over the FOV. Figure 3(c) represents the normalized average intensity as a function of focal position measured for a round-core fiber. We see that the average intensity significantly decreases towards the fiber edge.

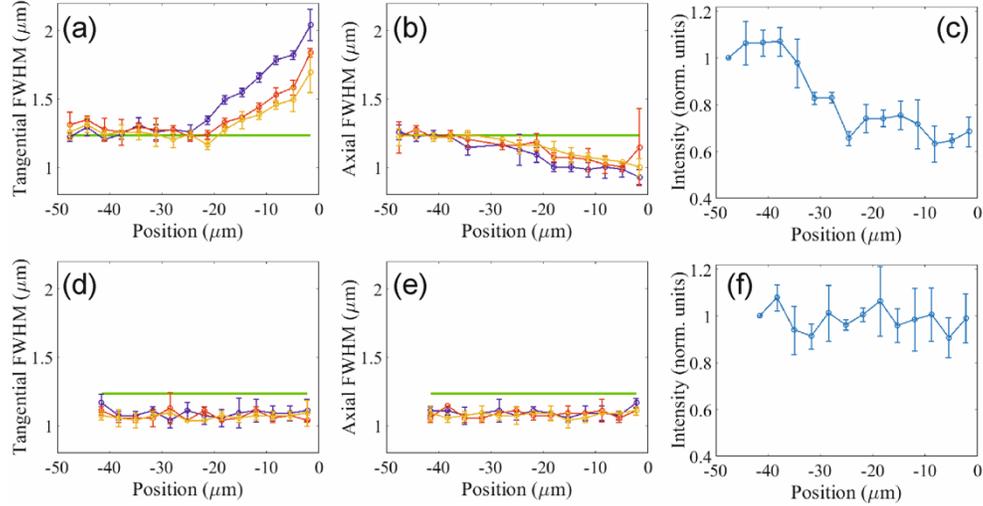

Fig. 3. (a,b) The FWHM in tangential (a) and axial (b) directions measured as function of the distance from the core edge (where the edge position is defined as 0) for a round-core fiber. (c) The normalized average intensity as a function of focal position for a round-core fiber. (d,e) The FWHM in tangential (d) and axial (e) directions measured as function of the distance from the core edge (where the edge position is defined as 0) for a square-core fiber. (f) The normalized average intensity as a function of focal position for a square-core fiber. The relative number of segments was kept constant for both fibers: $N_{seg}$ = 5% (blue), 14% (red), 22% (yellow) of the total number of modes supported by the MMF. The solid green line indicates the theoretical diffraction limit of the MMFs.

A similar analysis has been done for the square-core fiber. The FWHM focal dimensions in the tangential and axial directions experimentally measured as functions of a distance from the core edge (where the edge position is defined as 0) for a square-core fiber are presented in Fig. 3(d) and 3(e), respectively. Different colors represent the different number of DMD segments used for wavefront shaping: $N_{seg}$ = 256 (blue), 676 (red), 1024 (yellow). Please note that the relative number of segments was kept constant for both fibers: $N_{seg}$ = 5% (blue), 14% (red), 22% (yellow) of the total number of modes supported by the MMF. The solid green line indicates the FWHM that corresponds to the diffraction limit of the square-core MMF. Since the measured widths are slightly smaller than expected, we speculate that the NA of the square-core fiber is slightly higher than specified by the supplier. The normalized average intensity as a function of the focal position for a square-core fiber is presented in Fig. 3(f). We see that this intensity is more uniform than in the case of the round-core fiber.

Figures 3(d,e) clearly demonstrate that in the case of the square-core fiber the tangential and axial widths of the foci are largely independent of the position and of the total number of used segments. Even if we control only about 5% of the total number of modes we get a diffraction-limited focus at any point within a large FOV. To summarize, a square-core fiber provides diffraction-limited illumination without any noticeable aberration over the whole FOV, whereas large-FOV imaging through a round-core fiber is very limited.

*4.2 Light scrambling for imaging*

In this section, we experimentally investigate which core shape of an MMF provides the most spatially uniform distribution of speckled light at the output fiber facet. Homogeneous spatial distribution of speckles is very important for many applications such as laser speckle imaging [24], super-resolution compressive imaging [14] and nonlinear imaging methods based on time-domain wavefront shaping [12].

In our experiments, we use a single spatial mode at the input of the round-core and the square-core MMFs and measure the resulting speckle pattern with a camera, as presented in Fig. 1(b). We tune the wavelength of our laser source within 799-801 nm and average the result to get rid of any unwanted effects of the particular wavelength. We vary the position of the input mode with respect to the fiber center to be able to address input-output correlation effects. The results are presented in Fig. 4(a) for the round-core fiber and in Fig. 4(b) for a square-core fiber. Left to right the position on the input mode is shifted from the center toward the edge. The approximate position of the input focal spot is shown with the black cross mark.

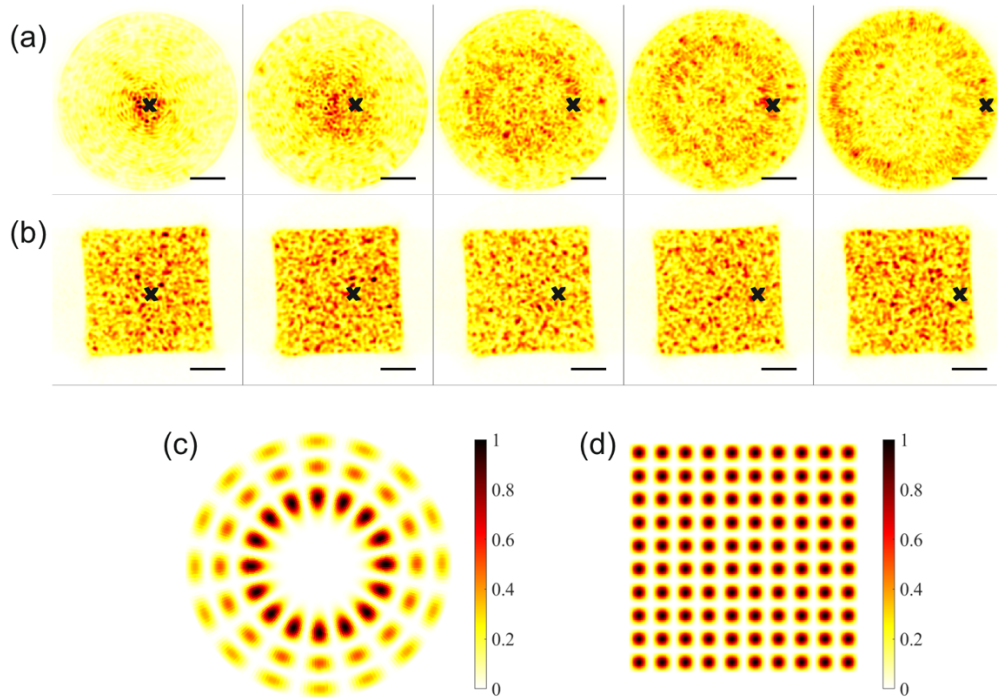

Fig. 4. (a,b) Speckle patterns on the MMF output facet with a round (a) and a square (b) core produced by a single-mode input and averaged over the wavelength range 795-805 nm. From left to right the position on the input is shifted from the center toward the edge as approximately shown with the black cross marks. (a, d) Examples of an eigenmode of a round-core (c) and square-core (d) fiber. Scale bars are 20 μm.

Figure 4(a) shows that the spatial intensity distribution at the output of the round-core fiber highly depends on the input position, even for a 1-meter-long bent probe. In contrast, the square-core multimode fiber provides a uniform speckle distribution on the fiber output facet, independent of the input location. This is a result of the different mode structures: Conservation of orbital angular momentum allows helical modes in a round-core fiber that tend to live at a certain radius, which depends highly on the input position, whereas the modes of a square-core fiber -that does not conserve angular momentum- fill out the core more uniformly. See Fig. 4(c, d) for an example of a high-order spatial mode of a round-core and a square-core fiber, respectively.

## 4.3 Light scrambling for spectrometry

The ability of multimode interference speckles to change their structure under tiny perturbations makes them a very sensitive tool which has been widely used for metrology and sensing applications. When the MMF is immobilized, the speckled output can be used to extract information about the illuminating light. One example is a speckle-based spectrometer, which retrieves the spectrum of light from the speckle pattern [25]. Here we investigate if the resolution of MMF-based spectrometers can be improved by utilizing square-core fibers. A simplified theoretical analysis of the spectrometer resolution shows that $\delta\lambda \sim \lambda^2 / [n\,L\,(\text{NA})^2]$, where $\lambda$ is the central wavelength, $n$ is the refractive index of the core and $L$ is the fiber length [6]. This simple expression captures the basic trends. However, it considers neither the complex amplitude profile of each mode nor the fact that only a fraction of the modes can be excited via a single-mode input. As a result, this simple equation can serve only as an upper limit and cannot be used for estimation of a real spectral resolution that is typically worse and can be affected by the fiber core shape.

In the final set of experiments, we couple light to an MMF and record the resulting speckle patterns for different wavelengths from 799.75 to 800.25 nm with a step of 6.5 pm (see Fig. 1(b)). The measurements were repeated for 5 different positions of the input beam with respect to the center of the fiber core: position #1 corresponds to the 'central' coupling where the input focus is aligned to the fiber center, position #5 corresponds to the 'edge coupling where the input focus is very close to the fiber core edge and positions #2-4 represent intermediate positions in the similar way as presented in Fig. 3(a,b). We calculate the Pearson correlation coefficients between the speckle pattern at 800 nm and all other speckle patterns. The correlation coefficient as a function of wavelength is presented in Fig. 5(a) for the round-core MMF and in Fig. 5(b) for the square-core MMF. Different colors correspond to different input positions. We see that for the square-core fiber the correlation function does not depend on the input position. In contrast, the input excitation position becomes critical for the conventional round-core fiber. The round fiber produces a broader correlation function with a significant background level. Figure 5(c) shows the width of the correlation function at the level of 0.5 as a function of the input position for a round-core (red) and a square-core (blue) MMF. The decorrelation width of the square-core fiber, in contrast, is independent of the coupling position.

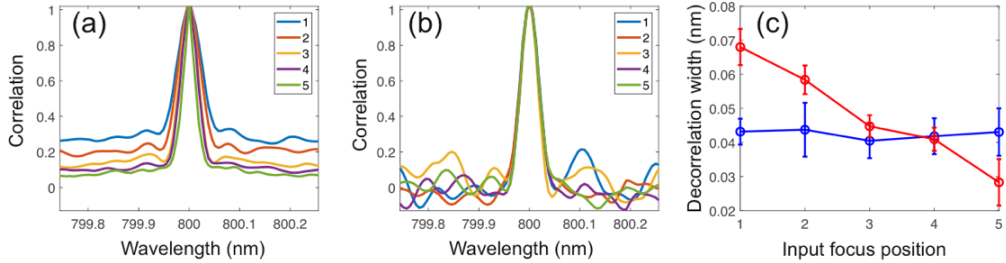

Fig. 5. (a,b) The Pearson correlation coefficient as a function of the wavelength for a round-core (a) and a square-core (b) MMF. Different colors correspond to different fiber inputs: position #1 corresponds to the 'central' position where the input focus is aligned to the fiber center, position #5 corresponds to the 'edge' position where the input focus is very close to the fiber core edge and positions #2-4 represent intermediate positions. (c) The width of the correlation function at the level of 0.5 as a function of the input position for a round-core (red) and a square-core (blue) MMF.

## 5. Conclusions

MMFs have already become a very important tool for remote sensing, imaging and spectroscopy. However, the search for the optimal fiber probe is still ongoing. Here we experimentally demonstrate that a square-core MMF doesn't suffer from aberrations as the traditional round-core MMF does. In contrast to a conventional round-core, wavefront shaping through a square-core MMF provides perfectly shaped foci at any position within a large field of view, even if a very low number of segments (5% of the total number of modes) is used for optimization. We also show that the decorrelation width of a square-core fiber is independent of the exact position of the input light, whereas a conventional round-core MMF produces a broader wavelength correlation function for light coupling in the fiber center. Finally, we demonstrate that a square-core shape provides a uniform distribution of the output intensity, free from any input-output correlations, showing great promise for imaging and spectroscopy applications.

**Funding** Funding is acknowledged from the Nederlandse Wetenschaps Organisatie (NWO) via QuantERA QUOMPLEX (Grant No. 680.91.037), NWA (Grant No. 40017607), and Veni (grant No. 15872).

**Acknowledgments**

**Disclosures** The authors declare no conflicts of interest.